\documentclass[amsmath,amssymb,aps,pre,preprint,groupedaddress]{revtex4-2}

\usepackage[dvipdfmx]{graphicx}
\usepackage{mathrsfs}
\usepackage{color}
\usepackage{amsthm}

\newtheorem*{conjecture*}{Conjecture}
\newtheorem*{approximation*}{Approximation}
\newtheorem*{theorem*}{Theorem}

\begin{document}


\title{Rate Distortion Theorem and the Multicritical Point of Spin Glass}



\author{Tatsuto Murayama}
\email[]{murayama@eng.u-toyama.ac.jp}
\affiliation{Graduate School of Science and Engineering,
University of Toyama,
3190 Gofuku, Toyama-shi, Toyama 930-8555, Japan}

\author{Asaki Saito}
\email[]{saito@fun.ac.jp}
\affiliation{Department of Complex and Intelligent Systems,
Future University Hakodate,
116-2 Kamedanakano-cho, Hakodate, Hokkaido 041-8655, Japan}

\author{Peter Davis}
\email[]{davis@telecognix.com}
\affiliation{Telecognix Corporation,
58-13 Yoshida Shimooji-cho, Sakyo-ku, Kyoto-shi, Kyoto 606-8314, Japan}


\date{\today}

\begin{abstract}
A spin system can be thought of as an information coding system
that transfers information of the interaction configuration
into information of the equilibrium state of the spin variables.
Hence it can be expected that the relations between the interaction configuration and equilibrium states
are consistent with the known laws of information theory.
We show that Shannon's rate-distortion theorem
can be used to obtain an universal constraint on neighboring spin correlations
for a broad range of Ising spin systems with two-body spin interactions.
Remarkably,
this constraint gives a bound for the multicritical point in the phase diagram, 
when a mean-field behavior for the neighboring spin pairs can be expected
in the paramagnetic phase.
\end{abstract}


\maketitle

Understanding the experimental observations of disordered materials has been a challenge to theoretical physicists.
This triggered the rise of a special area of statistical mechanics that deals with a variety of statistical models with frozen disorder,
where a series of mathematical techniques has become a common language for the systematic analysis~\cite{meza1987,meza2009}.
Moreover,
these techniques of statistical mechanics have been applied to the study of communication and information systems~\cite{rich2008,merhav2010statistical},
including noisy channel coding~\cite{sour1989,vicente1999finite,macris2007griffith},
recursive data compression~\cite{murayama2004thouless,ciliberti2005lossy,wainwright2010lossy},
CDMA multiuser detection~\cite{tanaka2002statistical,guo2005randomly,mimura2014generating},
modern cryptography~\cite{kabashima2000cryptographical},
and some combinatorial optimization problems and methods for them~\cite{wong1987graph,hukushima1996exchange,monasson1999determining}.
Overall,
the physicist's toolbox has successfully been applied to solve 
issues of information science; but {\it not} vice versa.
To our knowledge,
no classical theorem in information theory
has been
used to analyze
the physics of 
complex condensed matter such as spin glass.

This Rapid Communication shows that information theory can be effectively applied
to the analysis of spin glass systems.
In our scenario,
each of the equilibrium states of the Ising spins is regarded as one encoding of the interaction configuration~\cite{opper2001advanced,tala2003}.
This scenario enables us to apply
the Shannon rate-distortion theorem of information coding theory~\cite{cover2012elements},
which then allows us to develop a new method for
investigating fundamental restrictions on the phase diagram.
As a result, we obtain a previously unknown general bound for the location of
the multicritical point for Ising spin glasses,
where paramagnetic, ferromagnetic and spin glass phases merge~\cite{nishimori2001absence}.
Remarkably,
our argument is independent of detail structure of the lattice.
Numerical studies of
problems related to the location of the multicritical point
for specific lattice models
have been carried out by many physicists~\cite{le1988location,singh1991spin}.
However,
we still have little knowledge about these significant issues from a theoretical
point of view~\cite{honecker2001universality,parisi2014diluted}.

In our spin glass model,
we assign a binary spin $S_i=\pm 1$ to each site $i$
and the local energy
$-J_{ij}S_i S_j$
to a set of pairwise bonds $(i,j)$
with a binary interaction $J_{ij}=\pm 1$.
We investigate a class of Ising spin systems with the Hamiltonian
\begin{align}
\mathscr{H}\{ \mathcal{S} \}\{ \mathcal{J} \}
=
-\sum_{(i,j)}
J_{ij}S_i S_j \ ,
\label{eq:hamiltonian}
\end{align}
only assuming that the total number of the sites $i$
and the bonds $(i,j)$ are $N$ and $M$, respectively.
Specifically,
we do not restrict the range of the sum $(i,j)$
in (\ref{eq:hamiltonian}).
This sum could be over nearest neighbors,
or it might include farther pairs, etc.
Special features of each lattice will be reflected
only through the ratio $R=N/M$.
For the simplicity,
this work deals with
a Hamiltonian with two-body interactions
to elucidate the benefit of Shannon's
rate-distortion theorem,
although the same arguments apply to other multi-body spin systems.

Each $J_{ij}$ is supposed to be distributed independently according to
the common distribution
\begin{align*}
P(J_{ij})=p \delta(1,J_{ij})+(1-p)\delta(-1,J_{ij})
\end{align*}
for
a value of disorder parameter
$p$ in the interval $1/2 < p \le 1$.
Here $\delta$ denotes the Kronecker's delta function
and
the set of interaction coefficients
$\mathcal{J}=\{ J_{ij} \}$ is called the Bernoulli($p$)
random variables.
In general,
we write the inverse temperature as $\beta$
and then the phase diagram of the system can be depicted
in the space of disorder parameter $p$ and temperature $1/\beta$.
Now,
we consider the Nishimori temperature $1/\beta_p$ for the spin system,
defined to be
\begin{align*}
e^{2\beta_p}=\frac{p}{1-p} \ .
\end{align*}
Notice that the above equation specifies a line, the Nishimori line,
in the space of $p$ and $1/\beta$~\cite{supplement}.
It has been shown that the multicritical point can be always found
on this line.
And so,
we can specify the multicritical point by giving a value
for the disorder parameter $p$,
say $p_c$.
Moreover,
since spin glass phase does not exist on the Nishimori line,
the multicritical point can be characterized as a ferromagnetic transition
along the line~\cite{nish2001}.

\begin{figure}[tp]
\begin{center}
\includegraphics[width=\linewidth]{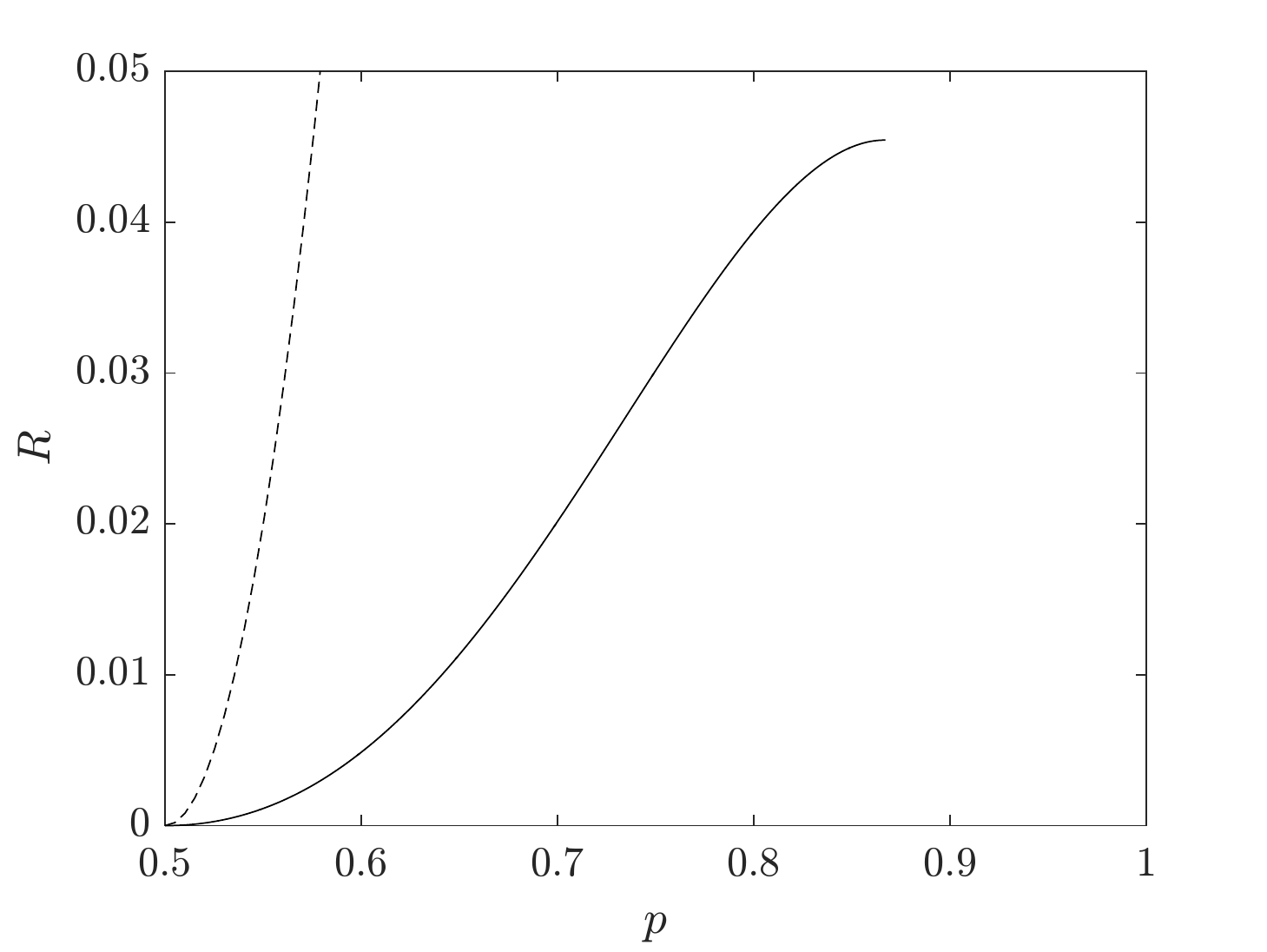}
\caption{
Theoretical bound for the multicritical point
in the phase diagram for disorder parameter $p$ and ratio $R=N/M$.
The solid line represents the upper bound $p^*$ for the transition point $p_c$ for a given $R$.
Shannon's rate-distortion theorem gives the constraint for
$R$ smaller than $0.0454$.
The dashed line represents the exact value of the transition point
$p_c$ for a family of the Bethe lattice spin glass.
}
\label{fig:bound}
\end{center}
\end{figure}

In this Rapid Communication,
we present a general bound
for the location of the multicritical point
of spin systems
on any lattice
with a Hamiltonian (\ref{eq:hamiltonian}).
Solid line in FIG.~\ref{fig:bound} shows the upper bound $p^*$
of $p_c$ for a given $R$,
only below which we find the multicritical point.
Notice that we can use Shannon's rate-distortion theorem 
to obtain this remarkable constraint when
a mean-field behavior can be expected in the paramagnetic phase.
More precisely, on the Nishimori line, we assume that
\begin{align}
    P(S_i,S_j | \mathcal{J}) \simeq \exp ( \beta_p J_{ij} S_i S_j )
    \label{eq:paramagnetic}
\end{align}
holds in the paramagnetic phase,
where the $P(S_i,S_j | \mathcal{J})$
denotes the joint distribution of $S_i$ and $S_j$ in the whole complex system
and the $\simeq$ means equality up to a normalization constant.
This implies that
the effect of the rest of the lattice on local marginals should not be dominant
and our potential target systems have a certain mean-field property
in the paramagnetic state, at least on the Nishimori line.
However,
we insist that
no further physical assumption is required to the Ising spin system.
As an example,
the dashed line in FIG.~\ref{fig:bound}
represents the exact value $p_c$
of the multicritical point for
a family of spin glass on a Bethe lattice~\cite{parisi2014diluted}.
Here all the bonds $(i,j)$ are chosen randomly to give a
diluted lattice with the fixed connectivity of $2/R$.
The standard cavity analysis shows that the relation (\ref{eq:paramagnetic}) holds
at any temperature
in the paramagnetic phase~\cite{mezard2001bethe,supplement}.
As is expected,
we can confirm that $p^*$ upper bounds $p_c$
for this specific model.

\begin{figure}[tp]
\begin{center}
\includegraphics[width=\linewidth]{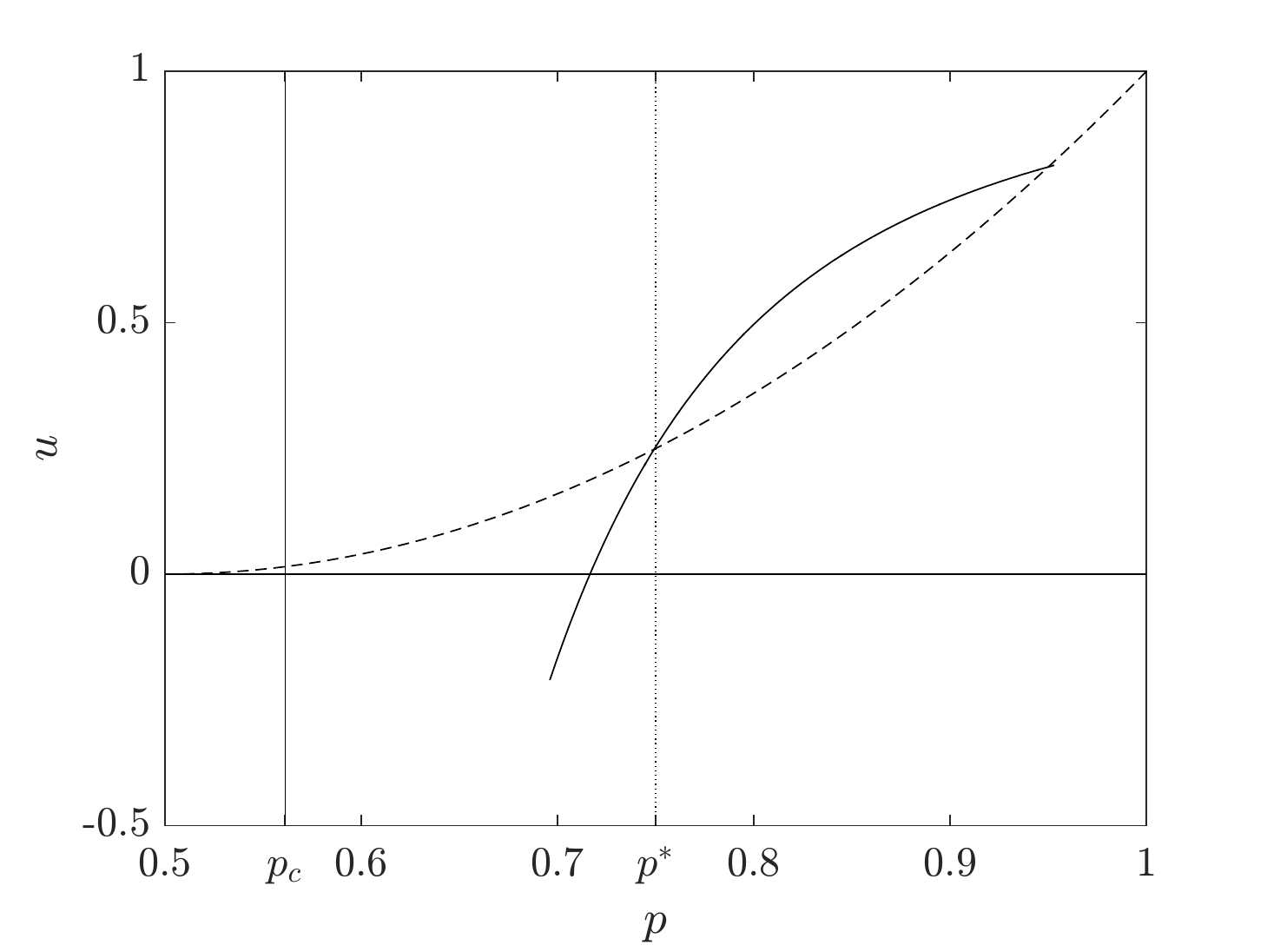}
\caption{
Theoretical constraints for local spin product $u$ as a function of disorder parameter $p$ for ratio $R=0.03$.
The solid curve
represents the universal lower bound of $u$ imposed by Shannon's rate-distortion theorem.
The dashed parabolic curve indicates the calculation of $u$
based on a mean-field assumption for neighboring spin marginals.
The paramagnetic solution contradicts our universal lower bound for $p$ greater than the intersection point $p^*=0.750$.
This means that for $p>p^*$ the paramagnetic solution is no more the stable solution,
implying that $p_c \leq p^*$.
The vertical solid line shows $p_c=0.561$ for a family of the Bethe lattice spin glass.}
\label{fig:theory}
\end{center}
\end{figure}

In the remainder of the work we will explain how this general bound can be obtained using the rate-distortion theorem.
For the reader's convenience,
we now outline the proof and then go into specific details afterwards.
We first define an average of local correlation functions
\begin{align*}
u=\Bigl[
\Big\langle
\frac{1}{M}\sum_{(i,j)} S_i S_j
\Bigr\rangle_\beta
\Bigr] \ ,
\end{align*}
where we assume that
$\langle \ \cdot \ \rangle_\beta$
represents the expectation value in the equilibrium state of the Hamiltonian (\ref{eq:hamiltonian})
at temperature $1/\beta$,
and suppose that a bracket $[ \ \cdot \ ]$
indicates
averaging over
an ensemble of configurations $\mathcal{J}$.
If the assumption (\ref{eq:paramagnetic}) holds within the paramagnetic phase,
we always get
$u=(2p-1)^2$
at $1/\beta_p$ for all $p < p_c$.
However,
if $R$ is small enough,
$u=(2p-1)^2$ derived from the paramagnetic assumption is smaller than the lower bound $u^*(p)$,
which is imposed by Shannon's rate-distortion theorem.
This implies that such $p$ for a given $R$ indicates the ferromagnetic state;
otherwise contradiction.
The infimum $p^*$ of such ferromagnetic $p$, therefore, gives an upper bound for the transition point $p_c$.
FIG.~\ref{fig:theory} illustrates a typical example with ratio $R=0.03$.

We first show that the local correlation function at $1/\beta_p$ is
\begin{align*}
u=(2p-1)^2 \ ,
\end{align*}
if the system is in the paramagnetic state.
Since the relation (\ref{eq:paramagnetic}) implies the explicit form
\begin{align*}
    P(S_i,S_j | \mathcal{J})=\frac{1}{4 \cosh \beta_p}\exp ( \beta_p J_{ij} S_i S_j ) \ ,
\end{align*}
it is an easy matter to check that
$\langle S_i S_j \rangle_{\beta_p}
=\tanh (\beta_p J_{ij})$
and averaging over $\mathcal{J}=\{ J_{ij} \}$ gives
$u=
[\tanh (\beta_p J_{ij})]
=(2p-1) \tanh \beta_p
=(2p-1)^2$.

Now,
we will explain how to obtain theoretical bound $u^*(p)$ for $u$
based on Shannon's rate-distortion theorem.
Let us first consider a virtual communication channel
where the interaction configuration sequence $\mathcal{J}=\{ J_{ij} \}$ is a
set of the Bernoulli($p$) random variables to be compressed, the set of spins $\mathcal{S}=\{ S_i \}$
is its compressed representation/codeword, and the spin products
$\hat{\mathcal{J}}=\{ S_i S_j \}$
are its reproduction at Nishimori temperature $1/\beta_p$.
This choice of communication channel is motivated by the fact that at the Nishimori temperature,
the Hamming distortion,
or the normalized Hamming distance,
$D=(1/M)\sum_{(i,j)} [\langle\delta (-1, J_{ij}S_i S_j)\rangle_{\beta_p}]$
between the $\mathcal{J}$ and its reproduction $\hat{\mathcal{J}}$ can be
easily
obtained as $D=1-p$~\cite{nish2001}.
This specific distortion measure defines the goodness of
$\hat{\mathcal{J}}$ as a representation of a set of given Bernoulli($p$) random variables $\mathcal{J}$.
The basic problem in Shannon's rate-distortion theory can then be stated as follows.
{\it What is the minimum description ratio $R=N/M$
required to achieve a given Hamming distortion $D$ between the two sequences?}
Shannon's rate-distortion theorem gives the lower bound, say $R_p(D)$,
as a function of the distortion measure $D$
for the theoretically achievable ratio $R=N/M$.
The ratio,
or rate,
$R_p(D)$ is called the rate-distortion function for the Bernoulli($p$)
random variables.
However,
the distortion $D=1-p$ only gives a trivial lower bound $R_p(D)=R_p(1-p)=0$
and results in no restrictions for this specific channel~\cite{cover2012elements}.

We thus introduce a coding `trick',
a set of the Bernoulli($\alpha$) random variables
$\tilde{\mathcal{J}}=\{ \tilde{J}_{ij} \}$ with $1/2 \le \alpha < p$,
which allows us to tighten the bound on $R=N/M$.
In the communication channel picture,
the manipulation of the Bernoulli($\alpha$) sequence $\tilde{\mathcal{J}}$
to get the sequence $\mathcal{J}$ corresponds to a preprocessing step in the encoding operation.
After we preprocess $\tilde{\mathcal{J}}$ to get $\mathcal{J}$,
the $\mathcal{J}$ is not Bernoulli($p$) assumed in the Nishimori's theory.
However, this difference becomes negligible when we take the large system limit of
$N \to \infty$.
As a result,
we can use the Nishimori's theory to calculate the Hamming distortion
between $\tilde{\mathcal{J}}$ and $\hat{\mathcal{J}}$,
which then offers a positive minimum ratio of $R=N/M$.
Since distortion $D$ redefined for the new pair depends on $p$ and $u$,
a positive bound on $R$ for the $D$, if any,
imposes a constraint on $u$ as a function of $p$ and $R$.
Hence,
we obtain the theoretical lower bound $u^*(p)$ on $u$ for a given ratio $R=N/M$.
Notice here that we require no physical assumptions such as (\ref{eq:paramagnetic})
in this argument.
In the following paragraphs we explain the essential details of this
universal analysis.

We first introduce a set
$\tilde{\mathcal{J}}$
of Bernoulli($\alpha$) random variables
for some $\alpha$ satisfying $1/2 \le \alpha < p$.
Define the set $T_a$ of all configurations with
relative frequency of $1$s equal to $a$.
For sufficiently large $M$, we can consider
$\tilde{\mathcal{J}} \in T_\alpha$
and
$\mathcal{J} \in T_p$,
respectively~\cite{csiszar2011information}.
So we suppose that any
$\tilde{\mathcal{J}}$
configuration can be switched to a
$\mathcal{J}$
configuration by flipping $(p-\alpha)M$ elements from $-1$ to $1$.
We consider the set of spin products
$\hat{\mathcal{J}}$
as an estimate of the original
$\tilde{\mathcal{J}}$.

Here
we evaluate
the normalized Hamming distance between the samples
$\tilde{\mathcal{J}}$ and
$\hat{\mathcal{J}}$,
i.e., $(1/M)\sum_{(i,j)}$
$\delta (-1, \tilde{J}_{ij} S_i S_j)$.
We first notice that the identity
$\tilde{J}_{ij} S_i S_j
=
J_{ij}\tilde{J}_{ij} \cdot J_{ij} S_i S_j$
leads to
\begin{align*}
\sum_{(i,j)} \delta (-1, \tilde{J}_{ij} S_i S_j)
\le
\sum_{(i,j)} \delta (-1, J_{ij}\tilde{J}_{ij})
+\sum_{(i,j)} \delta (-1, J_{ij}S_i S_j) \ .
\end{align*}
The equality holds if and only if there is no chance of getting
$J_{ij}\tilde{J}_{ij} = -1$ and $J_{ij}S_iS_j = -1$ simultaneously.
By definition,
the preprocessing gives
$\sum_{(i,j)} \delta (-1, J_{ij}\tilde{J}_{ij})=(p-\alpha)M$.
The second term on the right would be
\begin{align}
\Bigl[
\Big\langle \sum_{(i,j)} \delta(-1, J_{ij} S_i S_j)
\Bigr\rangle_{\beta_p}
\Bigr]
=(1-p)M \ ,
\label{eq:nishimori}
\end{align}
since the gauge theory tells us that the internal energy becomes
$[\langle \mathscr{H}\{ \mathcal{S} \}\{ \mathcal{J} \} \rangle_{\beta_p}]=-M \tanh \beta_p$
on the Nishimori line~\cite{supplement}.
Assume that the bracket $[ \ \cdot \ ]$
also indicates
averaging over
an ensemble of configurations $\tilde{\mathcal{J}}$
as well as $\mathcal{J}$.
Then we have
\begin{align}
\Bigl[
\Big\langle \sum_{(i,j)} \delta (-1, \tilde{J}_{ij} S_i S_j)
\Bigr\rangle_{\beta_p}
\Bigr]
\le
(1-\alpha)M \ .
\label{eq:upper}
\end{align}
To directly calculate the Hamming distance
between the samples
$\tilde{\mathcal{J}}$ and $\hat{\mathcal{J}}$
on the Nishimori line,
we introduce a pair of auxiliary variables
$Q_{1 \to -1}$
and
$Q_{-1 \to 1}$
defined to be
\begin{align*}
Q_{-1 \to 1} \ (1-p) M+Q_{1 \to -1} \ pM&=(1-p) M \ , \\
(1-Q_{-1 \to 1}) \ (1-p) M+Q_{1 \to -1} \ pM&=(1-q) M \ ,
\end{align*}
where $Q_{x \to y}$
is the empirical
probability of $S_i S_j=y$ when $J_{ij} = x$
and $q$ denotes a frequency of $1$s at the random variables $\hat{\mathcal{J}}$.
Notice that
the former equation just counts up every difference $J_{ij} \neq S_i S_j$,
while the latter indicates the total number of $S_i S_j=-1$
in the reconstruction.
By solving the two equations, we have
\begin{align*}
Q_{1 \to -1}
=
\frac{1-q}{2p} \ , \quad
Q_{-1 \to 1}
=
1-\frac{1-q}{2(1-p)} \ .
\end{align*}
It is easy to check that these formulas are well defined as
probabilities
in the interval $2p-1 \le q \le 1$ for a given $p \neq 1$.
Notice also that $u=2q-1$.
Then it follows that
\begin{align}
\Bigl[
\Big\langle \sum_{(i,j)} \delta (-1, \tilde{J}_{ij} S_i S_j)
\Bigr\rangle_{\beta_p}
\Bigr]
=
(1-\alpha)M-2Q_{1 \to -1}(p-\alpha)M
\label{eq:subtract}
\end{align}
(see \cite{supplement}).
In other words,
the normalized Hamming distance between
$\tilde{\mathcal{J}}$ and
$\hat{\mathcal{J}}$
on the Nishimori line
can be estimated by the formula
\begin{align*}
d_\alpha(p,q)=(1-\alpha)-2Q_{1 \to -1}(p-\alpha) \ ,
\end{align*}
which is non-negative for the relevant intervals.

Lastly,
it is possible to invoke Shannon's rate-distortion theorem
for the Bernoulli($\alpha$) random variables~\cite{cover2012elements}.
In the new communication channel picture with preprocessing,
we first write
$D=(1/M)\sum_{(i,j)} [\langle \delta (-1, \tilde{J}_{ij}S_i S_j) \rangle_{\beta_p}]$
and focus on the Hamming distortion
between the original $\tilde{\mathcal{J}}$ and its reproduction $\hat{\mathcal{J}}$.
Define the rate-distortion function
for the Bernoulli($\alpha$) random variables
as
\begin{align*}
R_\alpha(D)=H_2(\alpha)-H_2(D) \ ,
\end{align*}
where we denote
$H_2(\alpha)=-\alpha \log_2(\alpha)-(1-\alpha) \log_2(1-\alpha)$.
For the ratio $R=N/M$ and the
distortion $D$, the theorem states that
\begin{align*}
R_\alpha(D)<R \ .
\end{align*}
This inequality provides a bound on the compression ratio $R$,
dependent only on distortion $D$.
By letting $D=d_\alpha(p,q)$,
we can use the formula
$R_\alpha(p,q)
=
H_2(\alpha)-H_2(d_\alpha(p,q))$
to lower bound the ratio as
$R_\alpha(p,q) < R$
for every $\alpha$ in the relevant interval $1/2 \le \alpha < p$.
Now write
\begin{align*}
R^*(p,q)=\underset{1/2 \le \alpha < p}{\mathrm{sup}}
R_\alpha(p,q) \ .
\end{align*}
It is obvious that we can still lower bound $R$ as
\begin{align}
R^*(p,q)
\le
R \ .
\label{eq:inequality}
\end{align}
Notice also that the $R^*(p,q)$
is a non-increasing continuous function of $q$.
Suppose that the ratio $R=N/M$ is small enough to satisfy an inequality
$R <R^*(p,2p-1) $.
Here the $R^*(p,2p-1)$ is the largest value of
$R^*(p,q)$ for $q$ over the interval $2p-1 \le q \le 1$.
Since $d_\alpha(p,1)=1-\alpha$,
it is an easy matter to check that $R^*(p,1)=0$ for every $p$.
Then,
by the intermediate value theorem,
there exists a number
$q^*(p)$
in the closed interval
$2p-1 \le q \le 1$ such that
\begin{align}
R^*(p, q^*(p))=R \ .
\label{eq:equality}
\end{align}
We compare the formulas (\ref{eq:inequality}) and (\ref{eq:equality})
to conclude that
\begin{align*}
q^*(p) \le q \ ,
\end{align*}
i.e., the
$q^*(p)$
lower bounds $q$.

For the $R=N/M$ small enough,
we numerically examine the equation (\ref{eq:equality})
which implicitly determines
$q^*(p)$
for a given pair of $p$ and $R$.
Evaluation of the equation
$R^*(p,2p-1)=R$
shows that
there exists such a solution $q^*(p)$ for some $p$
for every $R$ smaller than $0.0541$.
Notice that the lower bound $q^*(p)$ for the Bernoulli parameter $q$
gives the lower bound $u^*(p)=2q^*(p)-1$ for local spin product $u$.
FIG.~\ref{fig:theory} compares this universal lower bound $u^*(p)$
with the preceding paramagnetic solution $u=(2p-1)^2$.
However, in this figure,
$u=(2p-1)^2$ violates our lower bound $u^*(p)$
for $p$ larger than the intersection point $p^*$.
Hence, the $p$ larger than $p^*$ implies the ferromagnetic phase,
in which the paramagnetic solution could break down.
In other words, the multicritical transition point $p_c$ should be smaller than the intersection point $p^*$.
For a given $R$,
this $p^*$ offers an upper bound for $p_c$ as is shown by the solid line in FIG.~\ref{fig:bound},
which is identified with $R^*(p,2p^2-2p+1)=R$.

In this Rapid Communication,
we considered the `$N$-bit' spin state of the Ising spin glass model
as compressed representations of a set of $M$ Bernoulli($p$) binary random variables
encoded in the interaction configuration.
We showed that the Shannon rate-distortion theorem,
which provides a bound on the compression ratio dependent only on distortion,
can give an upper bound $p^*$
for the location of the multicritical point $p_c$
for a sufficiently small
compression ratio $R=N/M$.
Remarkably, our argument is independent of detail structure of the lattice
and only requires a mean-field assumption for the joint marginals
of neighboring spins in the paramagnetic phase.
Results obtained here for a certain class of lattice models with two-body Ising spin interactions
will motivate applications of Shannon's rate-distortion theorem to other Ising spin systems.

\subsubsection*{Acknowledgments}

We would like to thank 
Federico Ricci-Tersenghi and 
Yoshiyuki Kabashima for useful discussions.
We also thank an anonymous reviewer for suggestions that improved the explanations.
This work was in part supported
by JSPS KAKENHI Grant Numbers JP16KK0005, JP17K00009.



\bibliography{APSrefs}

\providecommand{\noopsort}[1]{}\providecommand{\singleletter}[1]{#1}%
\begin{thebibliography}{29}%
\makeatletter
\providecommand \@ifxundefined [1]{%
 \@ifx{#1\undefined}
}%
\providecommand \@ifnum [1]{%
 \ifnum #1\expandafter \@firstoftwo
 \else \expandafter \@secondoftwo
 \fi
}%
\providecommand \@ifx [1]{%
 \ifx #1\expandafter \@firstoftwo
 \else \expandafter \@secondoftwo
 \fi
}%
\providecommand \natexlab [1]{#1}%
\providecommand \enquote  [1]{``#1''}%
\providecommand \bibnamefont  [1]{#1}%
\providecommand \bibfnamefont [1]{#1}%
\providecommand \citenamefont [1]{#1}%
\providecommand \href@noop [0]{\@secondoftwo}%
\providecommand \href [0]{\begingroup \@sanitize@url \@href}%
\providecommand \@href[1]{\@@startlink{#1}\@@href}%
\providecommand \@@href[1]{\endgroup#1\@@endlink}%
\providecommand \@sanitize@url [0]{\catcode `\\12\catcode `\$12\catcode
  `\&12\catcode `\#12\catcode `\^12\catcode `\_12\catcode `\%12\relax}%
\providecommand \@@startlink[1]{}%
\providecommand \@@endlink[0]{}%
\providecommand \url  [0]{\begingroup\@sanitize@url \@url }%
\providecommand \@url [1]{\endgroup\@href {#1}{\urlprefix }}%
\providecommand \urlprefix  [0]{URL }%
\providecommand \Eprint [0]{\href }%
\providecommand \doibase [0]{https://doi.org/}%
\providecommand \selectlanguage [0]{\@gobble}%
\providecommand \bibinfo  [0]{\@secondoftwo}%
\providecommand \bibfield  [0]{\@secondoftwo}%
\providecommand \translation [1]{[#1]}%
\providecommand \BibitemOpen [0]{}%
\providecommand \bibitemStop [0]{}%
\providecommand \bibitemNoStop [0]{.\EOS\space}%
\providecommand \EOS [0]{\spacefactor3000\relax}%
\providecommand \BibitemShut  [1]{\csname bibitem#1\endcsname}%
\let\auto@bib@innerbib\@empty
\bibitem [{\citenamefont {M{\'e}zard}\ \emph {et~al.}(1987)\citenamefont
  {M{\'e}zard}, \citenamefont {Parisi},\ and\ \citenamefont
  {Virasoro}}]{meza1987}%
  \BibitemOpen
  \bibfield  {author} {\bibinfo {author} {\bibfnamefont {M.}~\bibnamefont
  {M{\'e}zard}}, \bibinfo {author} {\bibfnamefont {G.}~\bibnamefont {Parisi}},\
  and\ \bibinfo {author} {\bibfnamefont {M.~A.}\ \bibnamefont {Virasoro}},\
  }\href@noop {} {\emph {\bibinfo {title} {Spin Glass Theory and Beyond}}}\
  (\bibinfo  {publisher} {World Scientific},\ \bibinfo {year}
  {1987})\BibitemShut {NoStop}%
\bibitem [{\citenamefont {M{\'e}zard}\ and\ \citenamefont
  {Montanari}(2009)}]{meza2009}%
  \BibitemOpen
  \bibfield  {author} {\bibinfo {author} {\bibfnamefont {M.}~\bibnamefont
  {M{\'e}zard}}\ and\ \bibinfo {author} {\bibfnamefont {A.}~\bibnamefont
  {Montanari}},\ }\href@noop {} {\emph {\bibinfo {title} {Information, Physics,
  and Computation}}}\ (\bibinfo  {publisher} {Oxford University Press},\
  \bibinfo {year} {2009})\BibitemShut {NoStop}%
\bibitem [{\citenamefont {Richardson}\ and\ \citenamefont
  {Urbanke}(2008)}]{rich2008}%
  \BibitemOpen
  \bibfield  {author} {\bibinfo {author} {\bibfnamefont {T.}~\bibnamefont
  {Richardson}}\ and\ \bibinfo {author} {\bibfnamefont {R.}~\bibnamefont
  {Urbanke}},\ }\href@noop {} {\emph {\bibinfo {title} {Modern Coding
  Theory}}}\ (\bibinfo  {publisher} {Cambridge University Press},\ \bibinfo
  {year} {2008})\BibitemShut {NoStop}%
\bibitem [{\citenamefont {Merhav}(2010)}]{merhav2010statistical}%
  \BibitemOpen
  \bibfield  {author} {\bibinfo {author} {\bibfnamefont {N.}~\bibnamefont
  {Merhav}},\ }\href@noop {} {\bibfield  {journal} {\bibinfo  {journal}
  {Foundations and Trends{\textregistered} in Communications and Information
  Theory}\ }\textbf {\bibinfo {volume} {6}},\ \bibinfo {pages} {1} (\bibinfo
  {year} {2010})}\BibitemShut {NoStop}%
\bibitem [{\citenamefont {Sourlas}(1989)}]{sour1989}%
  \BibitemOpen
  \bibfield  {author} {\bibinfo {author} {\bibfnamefont {N.}~\bibnamefont
  {Sourlas}},\ }\href@noop {} {\bibfield  {journal} {\bibinfo  {journal}
  {Nature}\ }\textbf {\bibinfo {volume} {339}},\ \bibinfo {pages} {693}
  (\bibinfo {year} {1989})}\BibitemShut {NoStop}%
\bibitem [{\citenamefont {Vicente}\ \emph {et~al.}(1999)\citenamefont
  {Vicente}, \citenamefont {Saad},\ and\ \citenamefont
  {Kabashima}}]{vicente1999finite}%
  \BibitemOpen
  \bibfield  {author} {\bibinfo {author} {\bibfnamefont {R.}~\bibnamefont
  {Vicente}}, \bibinfo {author} {\bibfnamefont {D.}~\bibnamefont {Saad}},\ and\
  \bibinfo {author} {\bibfnamefont {Y.}~\bibnamefont {Kabashima}},\ }\href@noop
  {} {\bibfield  {journal} {\bibinfo  {journal} {Physical Review E}\ }\textbf
  {\bibinfo {volume} {60}},\ \bibinfo {pages} {5352} (\bibinfo {year}
  {1999})}\BibitemShut {NoStop}%
\bibitem [{\citenamefont {Macris}(2007)}]{macris2007griffith}%
  \BibitemOpen
  \bibfield  {author} {\bibinfo {author} {\bibfnamefont {N.}~\bibnamefont
  {Macris}},\ }\href@noop {} {\bibfield  {journal} {\bibinfo  {journal} {IEEE
  Transactions on Information Theory}\ }\textbf {\bibinfo {volume} {53}},\
  \bibinfo {pages} {664} (\bibinfo {year} {2007})}\BibitemShut {NoStop}%
\bibitem [{\citenamefont {Murayama}(2004)}]{murayama2004thouless}%
  \BibitemOpen
  \bibfield  {author} {\bibinfo {author} {\bibfnamefont {T.}~\bibnamefont
  {Murayama}},\ }\href@noop {} {\bibfield  {journal} {\bibinfo  {journal}
  {Physical Review E}\ }\textbf {\bibinfo {volume} {69}},\ \bibinfo {pages}
  {035105} (\bibinfo {year} {2004})}\BibitemShut {NoStop}%
\bibitem [{\citenamefont {Ciliberti}\ \emph {et~al.}(2005)\citenamefont
  {Ciliberti}, \citenamefont {M{\'e}zard},\ and\ \citenamefont
  {Zecchina}}]{ciliberti2005lossy}%
  \BibitemOpen
  \bibfield  {author} {\bibinfo {author} {\bibfnamefont {S.}~\bibnamefont
  {Ciliberti}}, \bibinfo {author} {\bibfnamefont {M.}~\bibnamefont
  {M{\'e}zard}},\ and\ \bibinfo {author} {\bibfnamefont {R.}~\bibnamefont
  {Zecchina}},\ }\href@noop {} {\bibfield  {journal} {\bibinfo  {journal}
  {Physical Review Letters}\ }\textbf {\bibinfo {volume} {95}},\ \bibinfo
  {pages} {038701} (\bibinfo {year} {2005})}\BibitemShut {NoStop}%
\bibitem [{\citenamefont {Wainwright}\ \emph {et~al.}(2010)\citenamefont
  {Wainwright}, \citenamefont {Maneva},\ and\ \citenamefont
  {Martinian}}]{wainwright2010lossy}%
  \BibitemOpen
  \bibfield  {author} {\bibinfo {author} {\bibfnamefont {M.~J.}\ \bibnamefont
  {Wainwright}}, \bibinfo {author} {\bibfnamefont {E.}~\bibnamefont {Maneva}},\
  and\ \bibinfo {author} {\bibfnamefont {E.}~\bibnamefont {Martinian}},\
  }\href@noop {} {\bibfield  {journal} {\bibinfo  {journal} {IEEE Transactions
  on Information Theory}\ }\textbf {\bibinfo {volume} {56}},\ \bibinfo {pages}
  {135} (\bibinfo {year} {2010})}\BibitemShut {NoStop}%
\bibitem [{\citenamefont {Tanaka}(2002)}]{tanaka2002statistical}%
  \BibitemOpen
  \bibfield  {author} {\bibinfo {author} {\bibfnamefont {T.}~\bibnamefont
  {Tanaka}},\ }\href@noop {} {\bibfield  {journal} {\bibinfo  {journal} {IEEE
  Transactions on Information theory}\ }\textbf {\bibinfo {volume} {48}},\
  \bibinfo {pages} {2888} (\bibinfo {year} {2002})}\BibitemShut {NoStop}%
\bibitem [{\citenamefont {Guo}\ and\ \citenamefont
  {Verd{\'u}}(2005)}]{guo2005randomly}%
  \BibitemOpen
  \bibfield  {author} {\bibinfo {author} {\bibfnamefont {D.}~\bibnamefont
  {Guo}}\ and\ \bibinfo {author} {\bibfnamefont {S.}~\bibnamefont
  {Verd{\'u}}},\ }\href@noop {} {\bibfield  {journal} {\bibinfo  {journal}
  {IEEE Transactions on Information Theory}\ }\textbf {\bibinfo {volume}
  {51}},\ \bibinfo {pages} {1983} (\bibinfo {year} {2005})}\BibitemShut
  {NoStop}%
\bibitem [{\citenamefont {Mimura}\ and\ \citenamefont
  {Okada}(2014)}]{mimura2014generating}%
  \BibitemOpen
  \bibfield  {author} {\bibinfo {author} {\bibfnamefont {K.}~\bibnamefont
  {Mimura}}\ and\ \bibinfo {author} {\bibfnamefont {M.}~\bibnamefont {Okada}},\
  }\href@noop {} {\bibfield  {journal} {\bibinfo  {journal} {IEEE Transactions
  on Information Theory}\ }\textbf {\bibinfo {volume} {60}},\ \bibinfo {pages}
  {3645} (\bibinfo {year} {2014})}\BibitemShut {NoStop}%
\bibitem [{\citenamefont {Kabashima}\ \emph {et~al.}(2000)\citenamefont
  {Kabashima}, \citenamefont {Murayama},\ and\ \citenamefont
  {Saad}}]{kabashima2000cryptographical}%
  \BibitemOpen
  \bibfield  {author} {\bibinfo {author} {\bibfnamefont {Y.}~\bibnamefont
  {Kabashima}}, \bibinfo {author} {\bibfnamefont {T.}~\bibnamefont
  {Murayama}},\ and\ \bibinfo {author} {\bibfnamefont {D.}~\bibnamefont
  {Saad}},\ }\href@noop {} {\bibfield  {journal} {\bibinfo  {journal} {Physical
  Review Letters}\ }\textbf {\bibinfo {volume} {84}},\ \bibinfo {pages} {2030}
  (\bibinfo {year} {2000})}\BibitemShut {NoStop}%
\bibitem [{\citenamefont {Wong}\ and\ \citenamefont
  {Sherrington}(1987)}]{wong1987graph}%
  \BibitemOpen
  \bibfield  {author} {\bibinfo {author} {\bibfnamefont {K.~Y.~M.}\
  \bibnamefont {Wong}}\ and\ \bibinfo {author} {\bibfnamefont {D.}~\bibnamefont
  {Sherrington}},\ }\href@noop {} {\bibfield  {journal} {\bibinfo  {journal}
  {Journal of Physics A: Mathematical and General}\ }\textbf {\bibinfo {volume}
  {20}},\ \bibinfo {pages} {L793} (\bibinfo {year} {1987})}\BibitemShut
  {NoStop}%
\bibitem [{\citenamefont {Hukushima}\ and\ \citenamefont
  {Nemoto}(1996)}]{hukushima1996exchange}%
  \BibitemOpen
  \bibfield  {author} {\bibinfo {author} {\bibfnamefont {K.}~\bibnamefont
  {Hukushima}}\ and\ \bibinfo {author} {\bibfnamefont {K.}~\bibnamefont
  {Nemoto}},\ }\href@noop {} {\bibfield  {journal} {\bibinfo  {journal}
  {Journal of the Physical Society of Japan}\ }\textbf {\bibinfo {volume}
  {65}},\ \bibinfo {pages} {1604} (\bibinfo {year} {1996})}\BibitemShut
  {NoStop}%
\bibitem [{\citenamefont {Monasson}\ \emph {et~al.}(1999)\citenamefont
  {Monasson}, \citenamefont {Zecchina}, \citenamefont {Kirkpatrick},
  \citenamefont {Selman},\ and\ \citenamefont
  {Troyansky}}]{monasson1999determining}%
  \BibitemOpen
  \bibfield  {author} {\bibinfo {author} {\bibfnamefont {R.}~\bibnamefont
  {Monasson}}, \bibinfo {author} {\bibfnamefont {R.}~\bibnamefont {Zecchina}},
  \bibinfo {author} {\bibfnamefont {S.}~\bibnamefont {Kirkpatrick}}, \bibinfo
  {author} {\bibfnamefont {B.}~\bibnamefont {Selman}},\ and\ \bibinfo {author}
  {\bibfnamefont {L.}~\bibnamefont {Troyansky}},\ }\href@noop {} {\bibfield
  {journal} {\bibinfo  {journal} {Nature}\ }\textbf {\bibinfo {volume} {400}},\
  \bibinfo {pages} {133} (\bibinfo {year} {1999})}\BibitemShut {NoStop}%
\bibitem [{\citenamefont {Opper}\ and\ \citenamefont
  {Saad}(2001)}]{opper2001advanced}%
  \BibitemOpen
  \bibfield  {author} {\bibinfo {author} {\bibfnamefont {M.}~\bibnamefont
  {Opper}}\ and\ \bibinfo {author} {\bibfnamefont {D.}~\bibnamefont {Saad}},\
  }\href@noop {} {\emph {\bibinfo {title} {Advanced Mean Field Methods: Theory
  and Practice}}}\ (\bibinfo  {publisher} {MIT press},\ \bibinfo {year}
  {2001})\BibitemShut {NoStop}%
\bibitem [{\citenamefont {Talagrand}(2003)}]{tala2003}%
  \BibitemOpen
  \bibfield  {author} {\bibinfo {author} {\bibfnamefont {M.}~\bibnamefont
  {Talagrand}},\ }\href@noop {} {\emph {\bibinfo {title} {Spin Glasses: A
  Challenge for Mathematicians: Cavity and Mean Field Models}}}\ (\bibinfo
  {publisher} {Springer},\ \bibinfo {year} {2003})\BibitemShut {NoStop}%
\bibitem [{\citenamefont {Cover}\ and\ \citenamefont
  {Thomas}(2012)}]{cover2012elements}%
  \BibitemOpen
  \bibfield  {author} {\bibinfo {author} {\bibfnamefont {T.~M.}\ \bibnamefont
  {Cover}}\ and\ \bibinfo {author} {\bibfnamefont {J.~A.}\ \bibnamefont
  {Thomas}},\ }\href@noop {} {\emph {\bibinfo {title} {Elements of Information
  Theory}}}\ (\bibinfo  {publisher} {John Wiley \& Sons},\ \bibinfo {year}
  {2012})\BibitemShut {NoStop}%
\bibitem [{\citenamefont {Nishimori}\ and\ \citenamefont
  {Sherrington}(2001)}]{nishimori2001absence}%
  \BibitemOpen
  \bibfield  {author} {\bibinfo {author} {\bibfnamefont {H.}~\bibnamefont
  {Nishimori}}\ and\ \bibinfo {author} {\bibfnamefont {D.}~\bibnamefont
  {Sherrington}},\ }\href@noop {} {\bibfield  {journal} {\bibinfo  {journal}
  {AIP Conference Proceedings}\ }\textbf {\bibinfo {volume} {553}},\ \bibinfo
  {pages} {67} (\bibinfo {year} {2001})}\BibitemShut {NoStop}%
\bibitem [{\citenamefont {Doussal}\ and\ \citenamefont
  {Harris}(1988)}]{le1988location}%
  \BibitemOpen
  \bibfield  {author} {\bibinfo {author} {\bibfnamefont {P.~L.}\ \bibnamefont
  {Doussal}}\ and\ \bibinfo {author} {\bibfnamefont {A.~B.}\ \bibnamefont
  {Harris}},\ }\href@noop {} {\bibfield  {journal} {\bibinfo  {journal}
  {Physical Review Letters}\ }\textbf {\bibinfo {volume} {61}},\ \bibinfo
  {pages} {625} (\bibinfo {year} {1988})}\BibitemShut {NoStop}%
\bibitem [{\citenamefont {Singh}(1991)}]{singh1991spin}%
  \BibitemOpen
  \bibfield  {author} {\bibinfo {author} {\bibfnamefont {R.~R.~P.}\
  \bibnamefont {Singh}},\ }\href@noop {} {\bibfield  {journal} {\bibinfo
  {journal} {Physical Review Letters}\ }\textbf {\bibinfo {volume} {67}},\
  \bibinfo {pages} {899} (\bibinfo {year} {1991})}\BibitemShut {NoStop}%
\bibitem [{\citenamefont {Honecker}\ \emph {et~al.}(2001)\citenamefont
  {Honecker}, \citenamefont {Picco},\ and\ \citenamefont
  {Pujol}}]{honecker2001universality}%
  \BibitemOpen
  \bibfield  {author} {\bibinfo {author} {\bibfnamefont {A.}~\bibnamefont
  {Honecker}}, \bibinfo {author} {\bibfnamefont {M.}~\bibnamefont {Picco}},\
  and\ \bibinfo {author} {\bibfnamefont {P.}~\bibnamefont {Pujol}},\
  }\href@noop {} {\bibfield  {journal} {\bibinfo  {journal} {Physical Review
  Letters}\ }\textbf {\bibinfo {volume} {87}},\ \bibinfo {pages} {047201}
  (\bibinfo {year} {2001})}\BibitemShut {NoStop}%
\bibitem [{\citenamefont {Parisi}\ \emph {et~al.}(2014)\citenamefont {Parisi},
  \citenamefont {Ricci-Tersenghi},\ and\ \citenamefont
  {Rizzo}}]{parisi2014diluted}%
  \BibitemOpen
  \bibfield  {author} {\bibinfo {author} {\bibfnamefont {G.}~\bibnamefont
  {Parisi}}, \bibinfo {author} {\bibfnamefont {F.}~\bibnamefont
  {Ricci-Tersenghi}},\ and\ \bibinfo {author} {\bibfnamefont {T.}~\bibnamefont
  {Rizzo}},\ }\href@noop {} {\bibfield  {journal} {\bibinfo  {journal} {Journal
  of Statistical Mechanics: Theory and Experiment}\ }\textbf {\bibinfo {volume}
  {2014}},\ \bibinfo {pages} {P04013} (\bibinfo {year} {2014})}\BibitemShut
  {NoStop}%
\bibitem [{sup()}]{supplement}%
  \BibitemOpen
  \href@noop {} {}\bibinfo {note} {See Supplemental Material for some remarks
  and detailed calculations.}\BibitemShut {Stop}%
\bibitem [{\citenamefont {Nishimori}(2001)}]{nish2001}%
  \BibitemOpen
  \bibfield  {author} {\bibinfo {author} {\bibfnamefont {H.}~\bibnamefont
  {Nishimori}},\ }\href@noop {} {\emph {\bibinfo {title} {Statistical Physics
  of Spin Glasses and Information Processing: An Introduction}}}\ (\bibinfo
  {publisher} {Oxford University Press},\ \bibinfo {year} {2001})\BibitemShut
  {NoStop}%
\bibitem [{\citenamefont {M{\'e}zard}\ and\ \citenamefont
  {Parisi}(2001)}]{mezard2001bethe}%
  \BibitemOpen
  \bibfield  {author} {\bibinfo {author} {\bibfnamefont {M.}~\bibnamefont
  {M{\'e}zard}}\ and\ \bibinfo {author} {\bibfnamefont {G.}~\bibnamefont
  {Parisi}},\ }\href@noop {} {\bibfield  {journal} {\bibinfo  {journal} {The
  European Physical Journal B-Condensed Matter and Complex Systems}\ }\textbf
  {\bibinfo {volume} {20}},\ \bibinfo {pages} {217} (\bibinfo {year}
  {2001})}\BibitemShut {NoStop}%
\bibitem [{\citenamefont {Csiszar}\ and\ \citenamefont
  {K{\"o}rner}(2011)}]{csiszar2011information}%
  \BibitemOpen
  \bibfield  {author} {\bibinfo {author} {\bibfnamefont {I.}~\bibnamefont
  {Csiszar}}\ and\ \bibinfo {author} {\bibfnamefont {J.}~\bibnamefont
  {K{\"o}rner}},\ }\href@noop {} {\emph {\bibinfo {title} {Information theory:
  coding theorems for discrete memoryless systems}}}\ (\bibinfo  {publisher}
  {Cambridge University Press},\ \bibinfo {year} {2011})\BibitemShut {NoStop}%
\end{thebibliography}%


\providecommand{\noopsort}[1]{}\providecommand{\singleletter}[1]{#1}%
\begin{thebibliography}{0}%
\makeatletter
\providecommand \@ifxundefined [1]{%
 \@ifx{#1\undefined}
}%
\providecommand \@ifnum [1]{%
 \ifnum #1\expandafter \@firstoftwo
 \else \expandafter \@secondoftwo
 \fi
}%
\providecommand \@ifx [1]{%
 \ifx #1\expandafter \@firstoftwo
 \else \expandafter \@secondoftwo
 \fi
}%
\providecommand \natexlab [1]{#1}%
\providecommand \enquote  [1]{``#1''}%
\providecommand \bibnamefont  [1]{#1}%
\providecommand \bibfnamefont [1]{#1}%
\providecommand \citenamefont [1]{#1}%
\providecommand \href@noop [0]{\@secondoftwo}%
\providecommand \href [0]{\begingroup \@sanitize@url \@href}%
\providecommand \@href[1]{\@@startlink{#1}\@@href}%
\providecommand \@@href[1]{\endgroup#1\@@endlink}%
\providecommand \@sanitize@url [0]{\catcode `\\12\catcode `\$12\catcode
  `\&12\catcode `\#12\catcode `\^12\catcode `\_12\catcode `\%12\relax}%
\providecommand \@@startlink[1]{}%
\providecommand \@@endlink[0]{}%
\providecommand \url  [0]{\begingroup\@sanitize@url \@url }%
\providecommand \@url [1]{\endgroup\@href {#1}{\urlprefix }}%
\providecommand \urlprefix  [0]{URL }%
\providecommand \Eprint [0]{\href }%
\providecommand \doibase [0]{https://doi.org/}%
\providecommand \selectlanguage [0]{\@gobble}%
\providecommand \bibinfo  [0]{\@secondoftwo}%
\providecommand \bibfield  [0]{\@secondoftwo}%
\providecommand \translation [1]{[#1]}%
\providecommand \BibitemOpen [0]{}%
\providecommand \bibitemStop [0]{}%
\providecommand \bibitemNoStop [0]{.\EOS\space}%
\providecommand \EOS [0]{\spacefactor3000\relax}%
\providecommand \BibitemShut  [1]{\csname bibitem#1\endcsname}%
\let\auto@bib@innerbib\@empty
\end{thebibliography}%

\end{document}



\title{Supplemental Material: Rate Distortion Theorem and the Multicritical Point of Spin Glass}



\author{Tatsuto Murayama}
\email[]{murayama@eng.u-toyama.ac.jp}
\affiliation{Graduate School of Science and Engineering,
University of Toyama,
3190 Gofuku, Toyama-shi, Toyama 930-8555, Japan}

\author{Asaki Saito}
\email[]{saito@fun.ac.jp}
\affiliation{Department of Complex and Intelligent Systems,
Future University Hakodate,
116-2 Kamedanakano-cho, Hakodate, Hokkaido 041-8655, Japan}

\author{Peter Davis}
\email[]{davis@telecognix.com}
\affiliation{Telecognix Corporation,
58-13 Yoshida Shimooji-cho, Sakyo-ku, Kyoto-shi, Kyoto 606-8314, Japan}


\date{\today}



\maketitle

\noindent

In this brief document,
we give some supportive information to help the readers
grasp background knowledge and understand technical details about our
Rapid Communication.
Notice that a number for each formula in this document
is identical to the number we have already assigned in the associated work.

\begin{figure}[bp]
    \begin{center}
    \includegraphics[width=\linewidth]{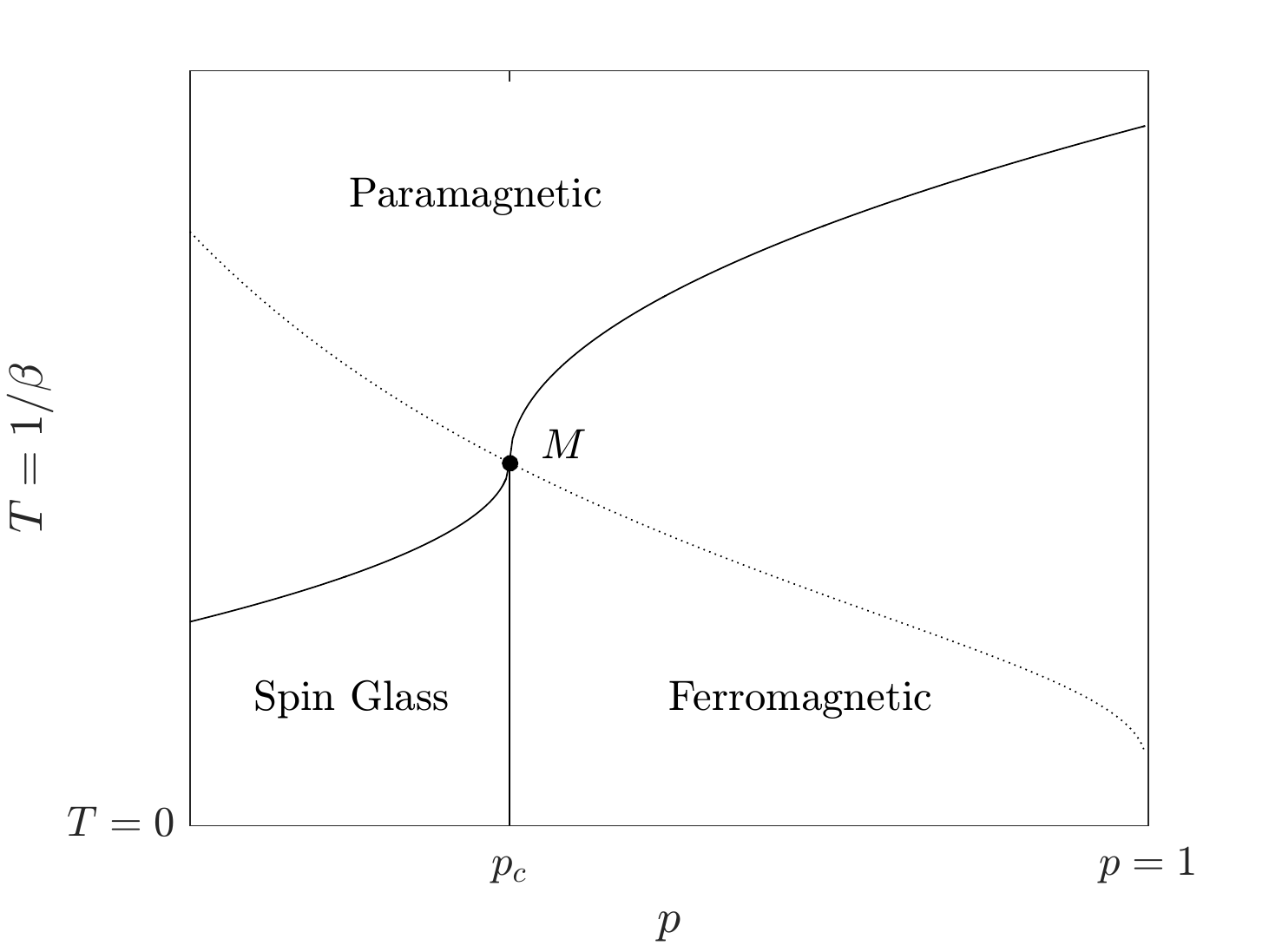}
    \caption{
        A schematic illustration of Nishimori line in the Ising spin glass models.
        The $M$ denotes the location of the multicritical point on the Nishimori line,
        where paramagnetic, ferromagnetic, and spin glass phases merge.
    }
    \label{fig:nishimori}
    \end{center}
\end{figure}

\noindent
{\bf Remarks on the Nishimori Line.}
We first illustrate a schematic picture
for the location of the multicritical point,
Nishimori line, and the three phase boundaries in the space of $p$ and $T=1/\beta$.
As is dipicted in FIG.~\ref{fig:nishimori}
the ferromagnetic phase transition along the Nishimori line
coincides with the multicritical point,
which enables us to identify the critical value $p_c$ for $p$
in the phase diagram.

\bigskip

\noindent
{\bf Remarks on the Bethe Lattice.}
We show that the Bethe lattice spin glass models satisfy the relation (2),
providing a pedagogical example.
We first notice that the Hamiltonian (1) gives
\begin{align*}
    \mathscr{H}\{ \mathcal{S} \}\{ \mathcal{J} \}
    &=
    -\sum_{(i,j)}
    J_{ij}S_i S_j
    =
    -\sum_{j \in \partial i}
    J_{ij}S_i S_j
    +\widehat{\mathscr{H}}_i \ .
\end{align*}
The $\partial i$ represents a collection of all sites $j$ connected to $i$
with an interaction $J_{ij}$.
The
$\widehat{\mathscr{H}}_i$
denotes the Hamiltonian of a smaller Ising spin system without the spin variable $S_i$.
Notice that the cavity field $h_{j \to i}$ is well defined on the Bethe lattice
through the equation
\begin{align*}
    \widehat{\mathscr{H}}_i
    =
    -\sum_{j \in \partial i} h_{j \to i} S_j \ .
\end{align*}
Then the effective Hamiltonian for the site $i$ and its vicinity is found to be
\begin{align*}
    \widehat{\mathscr{H}}
    =
    -\sum_{j \in \partial i}
    J_{ij}S_i S_j
    -\sum_{j \in \partial i} h_{j \to i} S_j \ .
\end{align*}
It follows that
\begin{align*}
    P(S_i | \mathcal{J})
    &=
    \frac{1}{\mathcal{Z}_1}
    \sum_{\{S_j \}_{j \in \partial i}}
    \exp\Bigl(-\beta \widehat{\mathscr{H}} \ \Bigr) \\
    &=
    \frac{1}{\mathcal{Z}_1}
    \sum_{\{S_j \}_{j \in \partial i}}
    \prod_{j \in \partial i}
    \exp
    \Bigl(
        \beta J_{ij}S_i S_j+\beta h_{j \to i} S_j
    \Bigr)
    &=
    \frac{1}{\mathcal{Z}_1}
    \prod_{j \in \partial i}
    \sum_{S_j=\pm 1}
    \exp
    \Bigl(
        \beta J_{ij}S_i S_j+\beta h_{j \to i} S_j
    \Bigr) \ .
\end{align*}
Here $\mathcal{Z}_1$ denotes the normalization constant.
Since $S_j=\pm 1$, the identity
\begin{align*}
    \exp
    \Bigl\{
        \beta S_j(J_{ij}S_i+h_{j \to i})
    \Bigr\}
    =
    2\cosh \{\beta ( J_{ij}S_i+h_{j \to i}) \}
    \frac{1+S_j \tanh \{\beta ( J_{ij}S_i+h_{j \to i}) \}}{2}
\end{align*}
gives
\begin{align*}
    \frac{1}{\mathcal{Z}_1}
    \prod_{j \in \partial i}
    \sum_{S_j=\pm 1}
    \exp
    \Bigl(
        \beta J_{ij}S_i S_j+\beta h_{j \to i} S_j
    \Bigr) 
    =
    \frac{1}{\mathcal{Z}_1}
    \prod_{j \in \partial i}
    \Bigl\{
    2\cosh \{\beta ( J_{ij}S_i+h_{j \to i}) \}
    \Bigr\} \ .
\end{align*}
Now we define the effective field
\begin{align*}
    \hat{h}_{j \to i}
    =
    \frac{1}{\beta}
    \tanh^{-1}
    \{
        \tanh (\beta J_{ij}) \tanh (\beta h_{j \to i})    
    \} \ .
\end{align*}
Together with another identity
\begin{align*}
    \cosh \{\beta ( J_{ij}S_i+h_{j \to i}) \}
    =
    2 \cosh (\beta J_{ij}) \cosh (\beta h_{j \to i})
    \frac{1+S_i \tanh (\beta J_{ij}) \tanh (\beta h_{j \to i})}{2} \ ,
\end{align*}
we have
\begin{align*}
    P(S_i | \mathcal{J})
    =
    \frac{1}{\mathcal{Z}_2}
    \exp \Bigl(\beta S_i \sum_{j \in \partial i} \hat{h}_{j \to i}\Bigr)
\end{align*}
with a new normalization constant $\mathcal{Z}_2$.
Since the magnetization at site $i$ is
\begin{align*}
    m_i
    =
    \frac{1}{\mathcal{Z}_2}
    \sum_{S_i=\pm 1}S_i
    \exp \Bigl(\beta S_i \sum_{j \in \partial i} \hat{h}_{j \to i}\Bigr) \ ,
\end{align*}
it is an easy matter to check that $h_{j \to i}=0$ for all $j$ gives $m_i=0$.
By letting $h_{j \to i}=0$,
the summation over the spin variables other than $S_i$ and $S_j$ results in
\begin{align*}
    P(S_i,S_j | \mathcal{J})
    =
    \frac{1}{\mathcal{Z}_3}\sum_{\{S_k \}_{k \in \partial i \backslash j}}
    \exp
    \Bigl(
        \beta \sum_{k \in \partial i}J_{ik}S_i S_k
    \Bigr) \ ,
\end{align*}
where $\mathcal{Z}_3$ denotes the normalization constant of the joint marginal distribution.
Finally, simple algebra gives
\begin{align*}
    P(S_i,S_j | \mathcal{J})=\frac{1}{4 \cosh \beta}\exp ( \beta J_{ij} S_i S_j ) \ .
\end{align*}
This indicates that the relation (2) holds at any temperature
in the paramagnetic phase for the Bethe lattice spin glass models and, hence,
the location of their multicritical points in the phase diagram
should be consistent with our results.

\bigskip

\noindent
{\bf Derivation of Equation (\ref{tag:3})}.
First,
the gauge theory tells us that the internal energy becomes
\begin{align*}
\frac{-1}{M}
[\langle \mathscr{H}\{ \mathcal{S} \}\{ \mathcal{J} \} \rangle_{\beta_p}]
=
\tanh \beta_p
=
2p-1
\end{align*}
at the Nishimori temperature $1/\beta_p$.
Notice that the above formula holds for any lattice.
Together with the definition of the Hamiltonian (1), i.e.,
\begin{align*}
\mathscr{H}\{ \mathcal{S} \}\{ \mathcal{J} \}
=
-\sum_{(i,j)}
J_{ij}S_i S_j
=
-\sum_{(i,j)} \delta (1, J_{ij}S_i S_j)
+\sum_{(i,j)} \delta (-1, J_{ij}S_i S_j) \ ,
\end{align*}
we get
\begin{align*}
\Bigl[
\Big\langle \sum_{(i,j)} \delta (1, J_{ij} S_i S_j)
\Bigr\rangle_{\beta_p}
\Bigr]
-
\Bigl[
\Big\langle \sum_{(i,j)} \delta (-1, J_{ij} S_i S_j)
\Bigr\rangle_{\beta_p}
\Bigr]
=
(2p-1)M \ .
\end{align*}
Notice also that
\begin{align*}
\Bigl[
\Big\langle \sum_{(i,j)} \delta (1, J_{ij} S_i S_j)
\Bigr\rangle_{\beta_p}
\Bigr]
+
\Bigl[
\Big\langle \sum_{(i,j)} \delta (-1, J_{ij} S_i S_j)
\Bigr\rangle_{\beta_p}
\Bigr]
=M
\end{align*}
holds. Then the conclusion follows, i.e.,
\begin{align}
    \Bigl[
    \Big\langle \sum_{(i,j)} \delta (-1, J_{ij} S_i S_j)
    \Bigr\rangle_{\beta_p}
    \Bigr]
    =
    (1-p)M \ . \tag{3} \label{tag:3}
\end{align}

\bigskip

\noindent
{\bf Derivation of Equation (\ref{tag:5})}.
We put
\begin{align*}
C &=\bigl\{
(i,j) \ | \ \tilde{J}_{ij}=-1 ~\text{or}~ 1
\bigr\} \ ,\\
D &=\bigl\{
(i,j) \ | \ \tilde{J}_{ij}=-1, \ J_{ij}=1, \ S_i S_j=-1
\bigr\} \ .
\end{align*}
Obviously, we have
\begin{align*}
\sum_{(i,j) \in D} \delta (-1, \tilde{J}_{ij} S_i S_j) &=0 \ ,\quad
\sum_{(i,j) \in D} \delta (-1, J_{ij}\tilde{J}_{ij}) &= |D| \ ,\quad
\sum_{(i,j) \in D} \delta (-1, J_{ij} S_i S_j) &= |D| \ ,
\end{align*}
where $|D|$ is the number of the elements in $D$.
As we discussed in the derivation of equation (4), we have
\[
\sum_{(i,j) \in C \setminus D} \delta (-1, \tilde{J}_{ij} S_i S_j)
=
\sum_{(i,j) \in C \setminus D} \delta (-1, J_{ij}\tilde{J}_{ij})
+
\sum_{(i,j) \in C \setminus D} \delta (-1, J_{ij} S_i S_j) \ .
\]
Thus, we have
\begin{align*}
\sum_{(i,j) \in C} \delta (-1, \tilde{J}_{ij} S_i S_j) &= \sum_{(i,j)
\in C \setminus D} \delta (-1, \tilde{J}_{ij} S_i S_j) + \sum_{(i,j) \in D}
\delta (-1, \tilde{J}_{ij} S_i S_j)\\
&= \sum_{(i,j) \in C \setminus D} \delta (-1, J_{ij}\tilde{J}_{ij})
+
\sum_{(i,j) \in C \setminus D} \delta (-1, J_{ij} S_i S_j) \\
&= \sum_{(i,j) \in C} \delta (-1, J_{ij}\tilde{J}_{ij})
+
\sum_{(i,j) \in C} \delta (-1, J_{ij} S_i S_j) -2 |D| \ .
\end{align*}
Taking the average, we obtain
\begin{align}
    \Bigl[
    \Big\langle \sum_{(i,j)} \delta (-1, \tilde{J}_{ij} S_i S_j)
    \Bigr\rangle_{\beta_p}
    \Bigr]
    =
    (1-\alpha)M-2Q_{1 \to -1}(p-\alpha)M \ . \tag{5} \label{tag:5}
\end{align}

\bibliography{APSrefs}